\documentclass[pra,aps,preprint,oneside,showpacs]{revtex4}
\usepackage{amsmath,amssymb,latexsym,graphicx}

\usepackage{graphicx}

\usepackage{bm}
\def\QQfnmark#1{\footnotemark}

\begin{document}

\vspace{-1cm}

\title{Configuration complexities of hydrogenic atoms}

\author{J.\ S.\ Dehesa}
\email{dehesa@ugr.es}
\affiliation{Departamento de F\'{\i}sica At\'{o}mica, Molecular y Nuclear and Instituto Carlos I de F\'{\i}sica Te\'orica y
Computacional, Universidad de Granada, 18071-Granada, Spain}

\author{S. L\'opez-Rosa}
\affiliation{Departamento de F\'{\i}sica At\'{o}mica, Molecular y Nuclear and Instituto Carlos I de F\'{\i}sica Te\'orica y
Computacional, Universidad de Granada, 18071-Granada, Spain}

\author{D. Manzano}
\affiliation{Departamento de F\'{\i}sica At\'{o}mica, Molecular y Nuclear and Instituto Carlos I de F\'{\i}sica Te\'orica y
Computacional, Universidad de Granada, 18071-Granada, Spain}

\date{\today}

\begin{abstract}

The Fisher-Shannon and Cramer-Rao information measures, and the LMC-like or shape complexity (i.e., the disequilibrium times the Shannon entropic power) of hydrogenic stationary states are investigated in both position and momentum spaces. First, it is shown that not only the Fisher information and the variance (then, the Cramer-Rao measure) but also the disequilibrium associated to the quantum-mechanical probability density can be explicitly expressed in terms of the three quantum numbers $(n,l,m)$ of the corresponding state. Second, the three composite measures mentioned above are analytically, numerically and physically discussed for both ground and excited states. It is observed, in particular, that these configuration complexities do not depend on the nuclear charge $Z$. Moreover, the Fisher-Shannon measure is shown to quadratically depend on the principal quantum number $n$. Finally, sharp upper bounds to the Fisher-Shannon measure and the shape complexity of a general hydrogenic orbital are given in terms of the quantum numbers.

\end{abstract}


\maketitle
\newpage

\section{Introduction}

A basic problem in information theory of natural systems is the identification of the proper quantifier(s) of their complexity or internal disorder at their physical states. Presently this remains open not only for a complicated system, like e.g. a nucleic acid (either DNA or its single-strand lackey, RNA) in its natural (decidedly non-crystalline) state, but also for the simplest quantum-mechanical realistic systems, including the hydrogenic atom. Indeed there does not yet exist any quantity to properly measure the rich variety of three-dimensional geometries of the hydrogenic orbitals, which are described by means of three integer numbers: the principal, orbital and magnetic or azimuthal quantum numbers usually denoted by n, l and m, respectively.

The root-mean-square or standard deviation does not measure the extent to which the electronic distribution is in fact concentrated, but rather the separation of the region(s) of concentration from a particular point of the distribution (the centroid or mean value), so that it is only useful for the nodeless ground state. In general, for excited states (whose probability densities are strongly oscillating) it is a misleading (and, at times, undefined) uncertainty measure. To take care of these defects, some information-theoretic quantities have been proposed: the Shannon entropic power \cite{SHA,STA} defined by
\begin{equation}\label{expS}
H \left[ \rho \right] = \exp \left\lbrace S \left[  \rho \right]\right\rbrace; \hspace*{1cm} \text{with} \hspace*{1cm} S \left[  \rho \right]= - \int \rho \left(\vec{r} \right)  \ln  \rho \left(\vec{r} \right)  d\vec{r},
\end{equation}
the averaging density or disequilibrium \cite{ONI,CAR,LOP,PIP,MAT,APL} defined by
\begin{equation}\label{dis}
\left\langle \rho \right\rangle = \int \rho^2 \left(\vec{r} \right)   d\vec{r},
\end{equation}
and the Fisher information \cite{FRI} defined by
\begin{equation}\label{fis}
I \left[ \rho \right] = \int \rho \left(\vec{r} \right)  \left[ \vec{\nabla} \ln \rho \left(\vec{r} \right) \right]^2   d\vec{r}.
\end{equation}

The two former quantities measure differently the total extent or spreading of the electronic distribution. Moreover, they have a global character because they are quadratic and logarithmic functionals of the associated probability density $\rho \left( \vec{r}\right)$. On the contrary, the Fisher information has a locality property because it is a gradient functional of the density, so that it measures the pointwise concentration of the electronic cloud and quantifies its gradient content, providing a quantitative estimation of the oscillatory character of the density. Moreover, the Fisher information measures the bias to particular points of the space, i.e. it gives a measure of the local disorder.

These three information-theoretic elements, often used as uncertainty measures, have shown (i) to be closely connected to various fundamental and/or experimentally measurable quantities (e.g., kinetic energy, ionization potential,..) (see e.g. \cite{SEA,A92}) and (ii) to exhibit the periodicity of the atomic shell structure (see e.g. \cite{ANG,JJS,PAN}). More recently, various composite information-theoretic measures have been introduced which have shown not only these properties but also other manifestations of the complexity of the atomic systems. Let us just mention the Fisher-Shannon measure defined by 
\begin{equation}\label{Cfs}
C_{FS} \left[ \rho \right]=  I \left[ \rho \right] \times  J \left[ \rho \right], \hspace*{1cm} \text{with} \hspace*{1cm} J  \left[ \rho \right]= \frac{1}{2 \pi e} \exp \left( 2 S \left[ \rho \right] /3 \right),
\end{equation}
the Cramer-Rao or Fisher-Heisenberg measure (see e.g. \cite{COV,DEM,ANG}) defined by
\begin{equation}\label{Ccr}
C_{CR} \left[ \rho \right]=  I \left[ \rho \right] \times  V \left[ \rho \right], \hspace*{1cm} \text{with} \hspace*{1cm} V  \left[ \rho \right]= \left\langle r^2 \right\rangle- \left\langle r \right\rangle^2,
\end{equation}
and the LMC-like or shape complexity \cite{LOP,GAR} defined by
\begin{equation}\label{Csc}
C_{SC} \left[ \rho \right]= \left\langle \rho \right\rangle  \times H \left[ \rho \right]
\end{equation}

They quantify different facets of the internal disorder of the system which are manifest in the diverse and complex three-dimensional geometries of its orbitals. The Fisher-Shannon measure grasps the oscillatory nature of the electronic probability cloud together with its total extent in the configuration space. The Cramer-Rao quantity takes also into account the gradient content but jointly with the electronic spreading around the centroid. The shape complexity measures the combined effect of the average height and the total spreading of the probability density; so, being insensitive to the electronic oscillations. This measure exhibits the important property of scale invariance, which the original LMC measure \cite{LOP} lacks, as it was first pointed out by Anteneodo and Plastino \cite{APL}

However, it has not yet been proved its usefulness to disentangle among the rich three-dimensional atomic geometries of any physical system, not even for the hydrogenic atom although some properties have been recently found \cite{SALO}. In this work we will investigate this issue by means of the three composite information-theoretic measures just mentioned for general hydrogenic orbitals in position and momentum spaces. Briefly, let us advance that here we find that the Fisher-Shannon measure turns out to be the most appropriate measure to describe the (intuitive) complexity of the three dimensional geometry of hydrogenic orbitals.

Nevertheless we should inmediately say that these three measures are complementary in the sense that, according to its composition, they grasp different facets of the internal disorder of the system which are manifest in the great diversity and complexity of configuration shapes of the probability density $\rho \left( \vec{r} \right)$ correspondig to its orbitals $(n,l,m)$. The Fisher-Shannon and Cramer-Rao measures have an ingredient of local character (namely, the Fisher information) and another one of global character (the modified Shannon entropic power in the Fisher-Shannon case and the variance in the Cramer-Rao case). The shape complexity is composed by two global ingredients: the disequilibrium and the Shannon entropic power; so, this quantity is not well prepared to grasp the oscillating nature of the hydrogenic orbitals but it takes into account the average height and the total extent of the electron distribution. The Fisher-Shannon measure appropriately describes the oscillating nature together with the total extent of the probability cloud of the orbital. The Cramer-Rao measure takes into account the gradient content jointly with the spreading of the probability density around its centroid.

The structure of the paper is the following. First, in Section II, the hydrogenic problem is briefly reviewed to fix notations and to gather the known results about the information-theoretic measures of the hydrogenic orbitals. In Section III, the disequilibrium of a generic orbital is explicitly calculated in terms of their quantum numbers. Then, in Section IV, the three composite measures mentioned above are discussed both numerically and analytically for the ground and excited hydrogenic states. In Section V, various sharp upper bounds for these composite measures are provided in terms of the three quantum numbers of the orbital. Finally, some conclusions are given.

\section{The hydrogenic problem: Information-theoretic measures}

In this Section we first describe the hydrogenic orbitals in the configuration space to fix notations; then we gather  some known results for various spreading measures (variance, Fisher information and Shannon entropy) of the system in terms of the quantum numbers $(n,l,m)$ of the orbital. Atomic units will be used throughout the paper.

The position hydrogenic orbitals (i.e., the solutions of the non-relativistic, time-independent Schr\"{o}dinger equation describing the quantum mechanics for the motion of an electron in the Coulomb field of a nucleus with charge $+Ze$) corresponding to stationary states of the hydrogenic system in the configuration space are characterized within the infinite-nuclear-mass approximation by the energetic eigenvalues
\begin{equation} \label{energia}
E= -\frac{Z^2}{2 n^2},\hspace{0.5cm}  \hspace{2cm} n=1,2,3,...,
\end{equation}
and the spatial eigenfunctions
\begin{equation}\label{FunOndaR}
\Psi_{n,l,m}(\vec{r})=R_{n,l}(r) Y_{l,m}(\Omega),
\end{equation}
where $n=1,2,...$, $l=0,1,...,n-1$ and $m=-l,-l+1,...,l-1,l$, and $r= \left| \vec{r} \right|$ and the solid angle $\Omega$ is defined by the angular coordinates $\left( \theta, \varphi \right)$. The radial eigenfunction, duly normalized to unity, is given by
\begin{equation}\label{Rnl}
R_{n,l}(r)= \frac{2 Z^{3/2}}{n^2} \left[\frac{\omega_{2l+1}(\tilde{r})}{\tilde{r}}\right]^{1/2}{\tilde{\cal{L}}}_{n-l-1}^{2l+1}(\tilde{r}),
\end{equation}
with $\tilde{r}=\frac{2 Zr}{n}$, and $\{{\tilde{\cal{L}}}_{k}^{\alpha} (x)\}$ denote the Laguerre polynomials orthonormal with respect to the weight function $\omega_\alpha (x)=x^\alpha e^{-x}$on the interval $\left[0,\infty\right)$; that is, they satisfy the orthogonality relation
\begin{equation}\label{OrtonormalidadL}
\int_{0}^{\infty} \omega_\alpha (x) {\tilde{\cal{L}}}_{n}^{\alpha} (x) {\tilde{\cal{L}}}_{m}^{\alpha} (x)= \delta_{nm}
\end{equation}

The angular eigenfuction $Y_{l,m}(\theta,\varphi)$ are the renowned spherical harmonics which describe the bulky shape of the system and are given by
\begin{equation}\label{Ylm}
Y_{l,m} (\theta,\varphi)= \frac{1}{\sqrt{2 \pi}} e^{i m \varphi}  {\tilde{\cal{C}}}_{l-m}^{m+1/2}\left(\cos \theta \right) \left(\sin \theta \right)^{m},
\end{equation}
where  $\{{{\tilde{\cal{C}}}_{k}^{\lambda}}(x)\}$ denotes the Gegenbauer or ultraspherical polynomials, wich are orthonormal with respect to the weight function $(1-x^2)^{\lambda-1/2}$ on the interval $\left[-1, +1\right]$. Then, the probability to find the electron between $\vec{r}$ and $\vec{r}+d\vec{r}$ is
\begin{equation}\nonumber
\rho \left( \vec{r}\right) d\vec{r}= \left|\Psi_{n,l,m}(\vec{r})  \right|^2 d\vec{r}= D_{n,l}(r) dr \times \Theta_{l,m}(\theta) d\theta d\varphi,
\end{equation}
where
\begin{equation}\label{DyTheta}
D_{n,l}=R_{n,l}^2 (r) r^2, \hspace*{1cm} \text{and} \hspace*{1cm} \Theta_{l,m}(\theta)=\left|Y_{l,n} (\theta,\varphi) \right|^2 \sin  \theta
\end{equation}
are the known radial and angular probability densities, respectively. So, the total probability density of the hydrogenic atom is given by
\begin{equation}\label{Rho}
\rho(\vec{r})= \frac{4 Z^3}{n^4} \frac{\omega_{2l+1}(\tilde{r})}{\tilde{r}} {\tilde{\cal{L}}}_{n-l-1}^{2l+1}(\tilde{r}) \left| Y_{l,m}(\theta,\varphi) \right|^2
\end{equation}

Let us now gather the known results for the following spreading measures of our system: the variance and the Fisher information. They have the values
\begin{equation}\label{Vrho}
V \left[\rho \right]=\frac{n^2 (n^2+2)-l^2 (l+1)^2}{4 Z^2}
\end{equation}
for the variance \cite{B57, H93}, and 
\begin{equation}\label{Irho}
I \left[\rho \right]=\frac{4 Z^2}{n^3} \left[ n -\left| m \right| \right]
\end{equation}
for the Fisher information \cite{R05, D05}.

The Shannon information of the hydrogenic atom $S \left[ \rho \right]$ is composed by the radial part given by
\begin{equation}\label{ShannonR}
S\left(R_{n,l}\right)= A_1(n,l)+\frac{1}{2 n} E_1\left( \tilde{{\cal{L}}}^{2l+1}_{n-l-1} \right)-3 \ln Z,
\end{equation}
with
\begin{equation}\nonumber
A_1 (n,l)= \ln \left( \frac{n^4}{4} \right)+\frac{3n^2-l (l+1)}{n}-2l \left[ \frac{2n-2l-1}{2n}+ \Psi (n+l+1) \right],
\end{equation}
where $\psi (x)= \Gamma{'}(x)/\Gamma(x)$ is the digamma function, and the angular part given by
\begin{equation}\label{ShannonY}
S\left(Y_{l,m}\right)= A_2(l,m)+ E \left( {\tilde{{\cal{C}}}}^{\left| m \right|+1/2}_{l- \left| m \right|} \right),
\end{equation}
\begin{equation}\nonumber
A_2 (l,m)=\ln \left(2^{2 \left| m \right|+1} \pi\right) -2 \left| m \right| \left[\psi (l+m+1)-\psi (l+1/2)-\frac{1}{2l+1} \right]
\end{equation}

Then, from Eqs. (\ref{ShannonR})-(\ref{ShannonY}), one has the value for the Shannon information of the state $(n,l,m)$:
\begin{eqnarray}\label{ShannonRho}
S \left[ \rho \right]&=& S\left(R_{n,l}\right)+ S\left(Y_{l,m}\right) \nonumber \\
&=& A(n,l,m)+\frac{1}{2 n} E_1\left( \tilde{{\cal{L}}}^{2l+1}_{n-l-1} \right)+E \left( {\tilde{{\cal{C}}}}^{\left| m \right|+1/2}_{l-\left| m \right|} \right)-3 \ln Z
\end{eqnarray}
with
\begin{eqnarray}\label{Ashannon}
A(n,l,m) & =& A_1(n,l)+A_2(l,m) \nonumber \\
& =& \ln \left(2^{2\left|m \right|-1} \pi n^4 \right)+\frac{3n^2-l(l+1)}{n} \nonumber \\
& \quad& - 2l \left[ \frac{2n-2l-1}{2n}+ \psi (n+l+1) \right]\nonumber \\
& \quad&-2 \left| m \right| \left[\psi (l+m+1)-\psi (l+1/2)-\frac{1}{2l+1} \right]
\end{eqnarray}

The symbols $E_i\left( {\tilde{y}}_{n} \right)$, $i=0$ and $1$, denote the following entropic integrals of the polynomials $\left\lbrace {\tilde{y}}_{n} \right\rbrace$ orthonormal with respect to the weight function $\omega (x)$ on $x$ $\epsilon \left[a,b \right]$
\begin{equation}\label{IntEntropicas0y1}
E_i \left( {\tilde{y}}_{n} \right) = \int_{a}^{b} x^i \omega (x) {\tilde{y}}_{n}^{2} (x) \ln {\tilde{y}}_{n}^{2} (x) dx,
\end{equation}
whose calculation is a difficult, not-yet-accomplished analytical task for polynomials of generic degree in spite of numerous efforts \cite{Y94,Y99,DMS,BUY}. As a particular case, let us mention that for the ground state $(n=1,l=m=0)$, Eqs. (\ref{Vrho}), (\ref{Irho}) and (\ref{ShannonRho})-(\ref{IntEntropicas0y1}) yield the following values
\begin{equation}\nonumber
V \left[ \rho_{g.s.} \right]=\frac{3}{4Z^2}, \hspace*{0.5cm} I \left[ \rho_{g.s.} \right]= 4Z^2 \hspace*{0.5cm} \text{and} \hspace*{0.5cm} S \left[ \rho_{g.s.} \right]=3+\ln \pi -3 \ln Z
\end{equation}
for the variance, Fisher information and Shannon entropy, respectively.

Let us now calculate the disequilibrium or averaging density $\left\langle \rho \right\rangle$ of the hydrogenic orbital $(n,l,m)$. From Eqs. (\ref{dis}) and (\ref{FunOndaR}) one has
\begin{eqnarray}\label{DeseqRho}
\left\langle \rho \right\rangle & =& \int_{\Re^3} \rho^2 \left( \vec{r} \right) d^3r= \int_{0}^{\infty} r^2 \left| R_{nl} (r) \right|^4 dr \times \int_{\Omega} \left|Y_{lm} \left( \Omega \right) \right|^4 d \Omega \nonumber \\
& \equiv& \left\langle \rho \right\rangle_R \times \left\langle \rho \right\rangle_Y
\end{eqnarray}

Let us begin with the calculation of the radial part $\left\langle \rho \right\rangle_R$. For purely mathematical convenience we use the notation $n_r=n-l-1$ and the change of variable $\tilde{r} =\frac{2 Z}{n}r $. Then one has
\begin{equation}\label{DeseqR}
\left\langle \rho \right\rangle_R =  \left( \frac{n}{2Z} \right)^3 \int_{0}^{\infty} \left| R_{nl} (\tilde{r}) \right|^4 {\tilde{r}}^2 d \tilde{r} = \frac{2 Z^3}{n^5} \left( \frac{n_r!}{(n+l)!} \right)^2 K \left(n_r,l \right),
\end{equation}
where $K \left(n_r,l \right)$ denotes the integral
\begin{eqnarray}\label{Kn_rl}
K \left(n_r,l \right)&=& \int_{0}^{\infty} e^{-2 \tilde{r}} {\tilde{r}}^{4l+2} \left[{\cal{L}}_{n_r}^{2l+1} (\tilde{r}) \right]^4 d \tilde{r} \\ \label{Kn_rl2}
&=& 2^{-4l-3} \left[ \frac{\Gamma \left(2l+n_r+2 \right)}{2^{2 n_r} n_r!} \right]^2 \sum_{k=0}^{n_r} \begin{pmatrix} 2n_r-2k \\[-1.5mm] n_r- k \end{pmatrix}^2 \frac{(2 k)! \Gamma \left(4l+2k+3 \right)}{\left(k!\right)^2 \Gamma^2 \left(2l+k+2 \right)}
\end{eqnarray}

For the second inequality, see Appendix. Then, the substitution of Eq. (\ref{Kn_rl2}) into Eq. (\ref{DeseqR}) yields the following value
\begin{equation}\label{DeseqR2}
\left\langle \rho \right\rangle_R = \frac{Z^3 2^{2-4n}}{n^5} \sum_{k=0}^{n_r} \begin{pmatrix} 2n_r-2k \\[-1.5mm] n_r- k \end{pmatrix}^2 \frac{(k+1)_k}{k!} \frac{ \Gamma \left(4l+2k+3 \right)}{ \Gamma^2 \left(2l+k+2 \right)}
\end{equation}
for the radial part of the disequilibrium. Remark that we have used the Pochhammer symbol $(x)_k= \Gamma(x+k)/\Gamma(x)$.

The angular contribution to the disequilibrium is
\begin{eqnarray}\nonumber
\left\langle \rho \right\rangle_Y &=& \int_{0}^{2 \pi} d \phi \int_{0}^{\pi} \sin \theta d \theta \left|Y_{lm} \left( \theta, \phi \right) \right|^4 \\ \label{DeseqY2}
&=& \sum_{l'=0}^{2l} \left( \frac{{\hat{l}}^2 \hat{l'}}{\sqrt{4\pi}} \right)^2 \begin{pmatrix} l & l & l' \\[-1.5mm] 0 & 0 & 0 \end{pmatrix}^2 \begin{pmatrix} l & l & l' \\[-1.5mm] m & m & -2m \end{pmatrix}^2
\end{eqnarray}
for the angular part of the disequilibrium.This expression is considerably much more transparent and simpler than its equivalent ${}_3F_{2} (1)$-form recently obtained \cite{DLY} by other means. See Appendix for further details.

Finally, the combination of Eqs. (\ref{DeseqRho}), (\ref{DeseqR2}) and (\ref{DeseqY2}) yields the value
\begin{equation}\label{DeseqRho2}
\left\langle \rho \right\rangle= Z^3 D (n,l,m)
\end{equation}
for the total disequilibrium of the hydrogenic orbital $\left(n,l,m \right)$, where $D(n,l,m)$ is given by
\begin{eqnarray}\label{Dnlm}
D(n,l,m)&=&\frac{(2l+1)^2}{2^{4n} \pi n^5} \sum_{k=0}^{n_r}  \begin{pmatrix} 2n_r-2k \\[-1.5mm] n_r-k \end{pmatrix}^2 \frac{\left( k+1 \right)_k}{k!} \frac{ \Gamma \left( 4l+2k+3\right)}{\Gamma^2 \left( 2l+k+2 \right)} \nonumber \\
& \quad& \times \sum_{l'=0}^{2l} (2l'+1) \begin{pmatrix} l & l & l' \\[-1.5mm] 0 & 0 & 0 \end{pmatrix}^2 \begin{pmatrix} l & l & l' \\[-1.5mm] m & m & -2m \end{pmatrix}^2,
\end{eqnarray}
where $n_r=n-l-1$. Note that for the ground state, the disequilibrium is $\left\langle \rho_{g.s.} \right\rangle = \frac{Z^3}{8\pi}$.

\section{Composite information-theoretic measures of hydrogenic orbitals}

Let us here discuss both analytical and numerically the three following composite information-theoretic measures of a general hydrogenic orbital with quantum numbers $(n,l,m)$: the Cramer-Rao or Fisher-Heisenberg and Fisher-Shannon measures and the shape complexity. Briefly, let us highlight in particular that these three quantities do not depend on the nuclear charge $Z$. Moreover, (a) the Cramer-Rao measure is given explicitly, (b) the Fisher-Shannon measure is shown to quadratically depend on the principal quantum number $n$, and (c) the shape complexity, which is a modified version of the LMC complexity \cite{LOP}, is carefully analyzed in terms of the quantum numbers. In this way we considerably extend the recent finding of Sa\~nudo and L\'opez-Ru\'iz \cite{SALO} relative to the fact that the Fisher-Shannon and shape complexities have their minimum values for the orbitals with the highest orbital momentum.

The Cramer-Rao measure is obtained in a straightforward manner from Eqs. (\ref{Ccr}), (\ref{Vrho}) and (\ref{Irho}), having the value
\begin{equation}\label{Ccr2}
C_{CR} \left[\rho \right] = \frac{n-\left| m \right| }{n^3} \left[ n^2 (n^2+2)-l^2 (l+1)^2 \right]
\end{equation}

Now, from Eqs. (\ref{Cfs}), (\ref{Irho}) and (\ref{ShannonR})-(\ref{Ashannon}) one has that the Fisher-Shannon measure has the value
\begin{equation}\label{Cfs2}
C_{FS} \left[\rho \right]=\frac{4 \left(n-\left| m \right| \right) }{n^3} \frac{1}{2 \pi e} e^{\frac{2}{3} B(n,l,m)},
\end{equation}
where
\begin{equation}\label{Bnlm}
B(n,l,m)= A(n,l,m)+\frac{1}{2n} E_1\left( \tilde{{\cal{L}}}^{2l+1}_{n-l-1} \right)+E \left( {\tilde{{\cal{C}}}}^{\left| m \right|+1/2}_{l-\left| m \right|} \right)
\end{equation}

The symbols $E_i \left({\tilde{y}}_n \right)$ denote the entropic integrals given by Eq. (\ref{IntEntropicas0y1}). Similarly, taking into account Eqs. (\ref{expS}), (\ref{Csc}), (\ref{ShannonRho})-(\ref{Ashannon}) and (\ref{DeseqRho2}) one has the value
\begin{equation}\label{Csc2}
C_{SC}  \left[\rho \right] = D(n,l,m) e^{B(n,l,m)},
\end{equation}
where the explicit expression of $D(n,l,m)$ is given by Eq. (\ref{Dnlm}). In particular, for the ground state we have the values
\begin{equation}\nonumber
C_{CR} \left[\rho_{g.s.} \right] =3, \hspace*{1cm}  C_{FS} \left[\rho_{g.s.} \right] =\frac{2 e}{\pi^{1/3}}, \hspace*{1cm} \text{and} \hspace*{1cm}  C_{SC} \left[\rho_{g.s.} \right] = \frac{e^3}{8}
\end{equation}
for the three composite information-theoretic measures mentioned above. Let us highlight from Eqs. (\ref{Cfs2})-(\ref{Csc2}) that the three composite information-theoretic measures do not depend on the nuclear charge. Moreover, it is known that $C_{FS} \left[ \rho \right] \geq 3$ for all three-dimensional densities \cite{STA, COV} but also $C_{CR} \left[ \rho \right] \geq 3$ for any hydrogenic orbital as one can easily show from Eq. (\ref{Ccr2}).

Let us now discuss numerically the Fisher-Shannon, Cramer-Rao and shape complexity measures of hydrogenic atoms for various specific orbitals in terms of their corresponding quantum numbers $(n,l,m)$. To make possible the mutual comparison among these measures and to avoid problems with physical dimensions, we study the dependence of the ratio between the measures $C \left[\rho_{n,l,m} \right] \equiv C (n,l,m) $ of the orbital we are interested in and the corresponding measure $C \left[\rho_{1,0,0} \right] \equiv C (1,0,0) \equiv C (g.s.)$ of the ground state, that is:
\begin{equation}\nonumber
\zeta (n,l,m) := \frac{C(n,l,m)}{C(1,0,0)},
\end{equation}
on the three quantum numbers. The results are shown in Figures 1, 2 and 3, where the relative values of the three composite information-theoretic measures are plotted in terms of $n$, $m$ and $l$, respectively. More specifically, in \textbf{Figure 1}, we have given the three measures for various ns-states (i.e., with $l=m=0$). Therein, we observe that (a) the Fisher-Shannon and Cramer-Rao measures have an increasing parabolic behaviour when $n$ is increasing while the shape complexity is relatively constant, and (b) the following inequalities

\begin{figure}[h!]
\includegraphics[scale=0.4,angle=270]{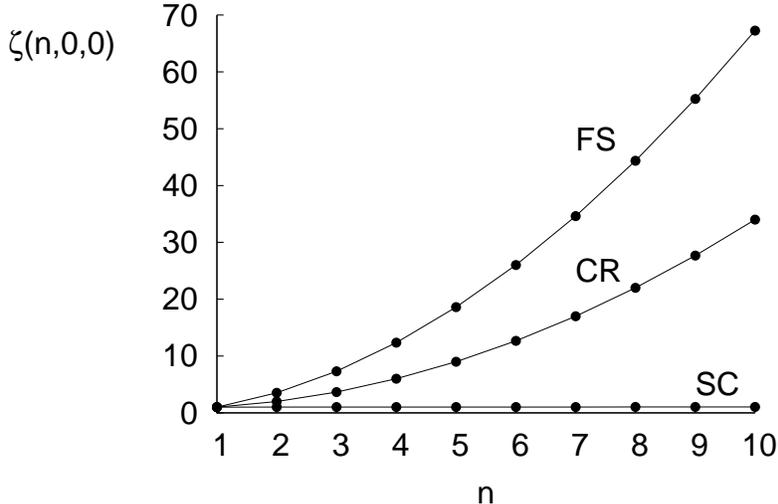}
\caption{Relative Fisher-Shannon measure $\zeta_{FS} (n,0,0)$, Cramer-Rao measure $\zeta_{CR} (n,0,0)$  and shape complexity $\zeta_{SC} (n,0,0)$ of the ten lowest hydrogenic states s as a function of $n$. See text.}
\end{figure}

\begin{equation}\nonumber
 \zeta_{FS}(n,0,0) > \zeta_{CR}(n,0,0) > \zeta_{SC} (n,0,0)
\end{equation}
are fulfilled for fixed $n$. Similar characteristics are shown by states $(n,l,m)$ other than $(n,0,0)$. Both to understand this behaviour and to gain a deeper insight into the internal complexity of the hydrogenic atom which is manifest in the three-dimensional geometry of its configuration orbitals (and so, in the spatial charge distribution density of the atom at different energies), we have drawn the radial $D_{n,l}=R_{n,l}^2(r) r^2$ and angular $\Theta_{l,m} \left( \theta \right) = \left|Y_{l,m} \left(\theta,\varphi \right) \right|^2 \sin \theta$ densities (see Eq. (\ref{DyTheta})) in \textbf{Figures 4} and \textbf{5}, respectively, for the three lowest energetic levels of hydrogen.

\begin{figure}[h!]
\includegraphics[scale=0.5]{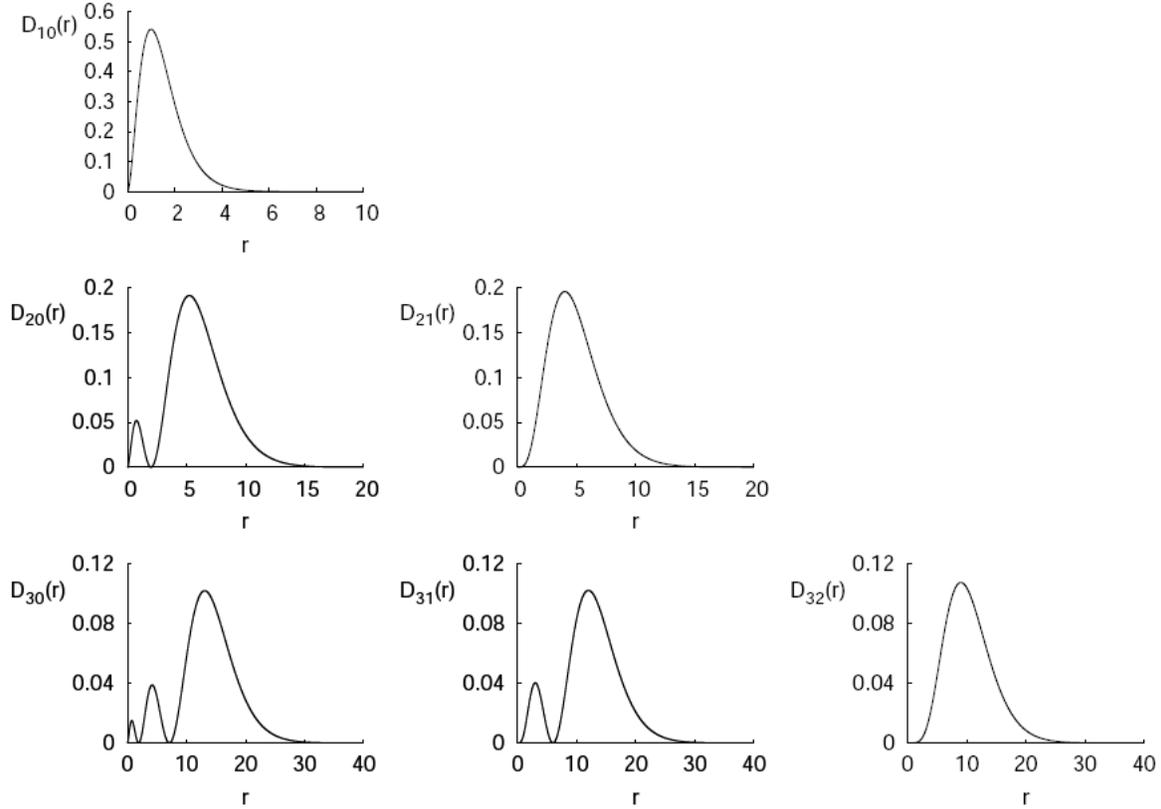}
\caption{Radial distribution $D_{n,l}= R_{n,l}^2(r)r^2$ of all the electronic orbitals corresponding to the three lowest energy levels of hydrogen. Atomic units have been used.}
\end{figure}

From Figure 4 we realize that when $n$ is increasing and l is fixed, both the oscillatory character (so, the gradient content and its associated Fisher information) and the spreading (so, the Shannon entropic power) of the radial density certainly grow while its variance hardly does so and the average height (which controls $\left\langle \rho \right\rangle$) clearly decreases. Taking into account these radial observations and the graph of $\Theta_{0,0} (\theta)$ at the top line of Figure 5, we can understand the parabolic growth of the Fisher-Shannon and Cramer-Rao measures as well as the lower value and relative constancy of the shape complexity for ns-states shown in Figure 1 when $n$ is increasing. In fact, the gradient content (mainly because of its radial contribution) and the spreading of the radial density of these states contribute constructively to the Fisher-Shannon measure of hydrogen, while the spreading and the average height almost cancel one to another, making the shape complexity to have a very small, almost constant value; we should say, for completeness, that $\zeta_{SC} (n,0,0)$ increases from $1$ to $1.04$ when $n$ varies from $1$ to $10$. In the Cramer-Rao case, the parabolic growth is practically only due to the increasing behaviour of the gradient content, so to its Fisher information ingredient.

\begin{figure}[h!]
\includegraphics[scale=0.5]{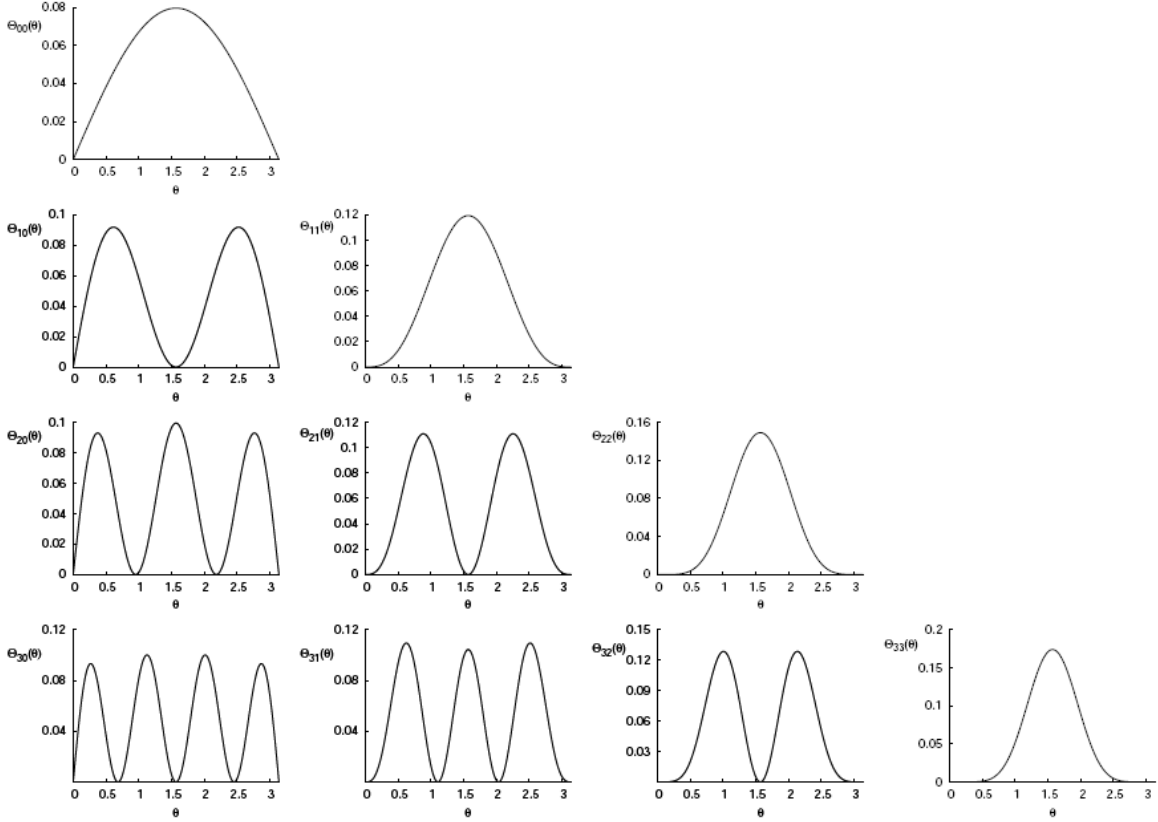}
\caption{Angular distribution $\Theta_{l,m} \left(\theta \right)= \left|Y_{l,m} \left(\theta, \varphi \right)\right|^2 \sin \theta$ of all electronic orbitals corresponding to the four lowest lying energy levels of hydrogen. Atomic units have been used.}
\end{figure}

Let us now explain and understand the linear decreasing behaviour of the Fisher-Shannon and Cramer-Rao measures as well as the pratical constancy of the shape complexity for the hydrogen orbital $(n=20,l=17,m)$ when $\left| m \right|$ is increasing, as shown in \textbf{Figure 2}. These phenomena purely depend on the angular contribution due to the analytical form of the angular density $\Theta_{17,m} (\theta)$ since the radial contribution (i.e. that due to the radial density $R_{n,l}(r)$) is constant when $m$ varies. A straightforward extrapolation of the graphs corresponding to the angular densities $\Theta_{l,m} (\theta)$ contained in Figure 5, shows that when $l$ is fixed and $\left| m \right|$ is increasing, both the gradient content and spreading of this density decrease while the average height and the probability concentration around its centroid are apparently constant. Therefore, the Fisher-Shannon and Cramer-Rao have a similar decreasing behaviour as shown in Figure 2 although with a stronger rate in the former case, because its two ingredients (Fisher information and Shannon entropic power) contribute constructively while in the Cramer-Rao case, one of the ingredients (namely, the variance) does not contribute at all. Keep in mind, by the way, that the relations (\ref{Vrho}) and (\ref{Irho}) show that the total variance does not depend on $m$  and the Fisher information linearly decreases when $\left| m \right|$ is increasing, respectively. On the other hand, Figure 5 shows that the angular average height increases while the spreading decreases so that the overall combined contribution of these two ingredients to the shape complexity is relatively constant and very small when $\left| m \right|$ varies; in fact, $\zeta_{SC} (20,17,m)$ parabolically decreases from $1$ to $0.6$ when $\left| m \right|$ varies from $0$ to $17$.

\begin{figure}[h!]
\includegraphics[scale=0.4,angle=270]{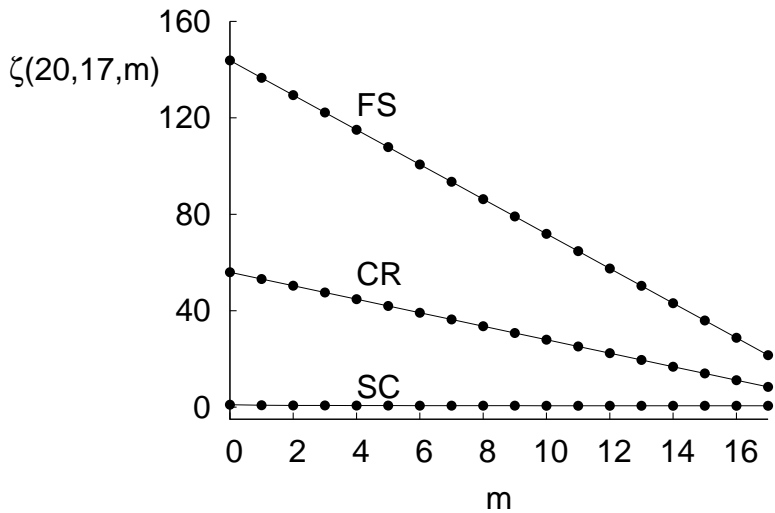}
\caption{Relative Fisher-Shannon measure $\zeta_{FS} (20,17,m)$, Cramer-Rao measure $\zeta_{CR} (20,17,m)$  and shape complexity $\zeta_{SC} (20,17,m)$ of the manifold of hydrogenic levels with $n=20$ and $l=17$ as a function of the magnetic quantum number $m$. See text.}
\end{figure}

In \textbf{Figure 3} it is shown that the Fisher-Shannon and Cramer-Rao measures have a concave decreasing form and the shape complexity turns out to be comparatively constant for the orbital $(n=20,l,m=1)$ when the orbital quantum number $l$ varies. We can undestand these phenomena by taking into account the graphs, duly extrapolated, of the files of Figure 4 and the columns of Figure 5 where the radial density for fixed $n$ and the angular density for fixed $m$ are shown. Herein we realize that when $l$ is increasing, (a) the radial gradient content decreases while the corresponding angular quantity increases, so that the gradient content of the total density $\rho \left(\vec{r} \right)$ does not depend on $l$ in accordance to its Fisher information as given by Eq. (\ref{Irho}); (b) the radial and angular spreadings have decreasing and constant behaviours, respectively, so that the overall effect is that the Shannon entropic power of the total density $\rho \left(\vec{r} \right)$ increases, (c) both the radial and the angular average height increase, so that the total averaging density $\rho \left(\vec{r} \right)$ increases, and a similar phenomenon occurs with the concentration of the radial and angular probability clouds around their respective mean value, so that the total variance $V \left[ \rho \right]$ decreases very fast (as Eq. (\ref{Vrho}) analytically shows). Taking into account these observations into the relations (\ref{Cfs}), (\ref{Ccr}) and (\ref{Csc}) which define the three composite information-theoretic measures under consideration, we can immediatly explain the decreasing dependence of the Fisher-Shannon and Cramer-Rao measures on the orbital quantum number as well as the relative constancy of the shape complexity, as illustrated in Figure 3; in fact, $\zeta_{SC} (20,l,1)$ also decreases but within the small interval $(1,0.76)$ when $l$ goes from $0$ to $19$.

\begin{figure}[h!]
\includegraphics[scale=0.4,angle=270]{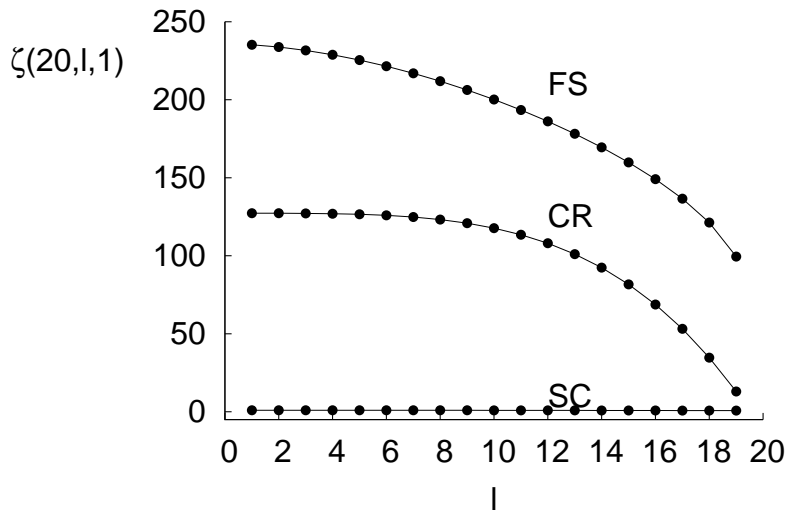}
\caption{Relative Fisher-Shannon measure $\zeta_{FS} (20,l,1)$, Cramer-Rao measure $\zeta_{CR}(20,l,1)$  and shape complexity $\zeta_{SC} (20,l,1)$ of the hydrogenic states with $n=20$ and $m=1$ as a function of the orbital quantum number $l$. See text.}
\end{figure}

Finally, for completeness, we have numerically studied the dependence of the Fisher-Shannon measure on the principal quantum number $n$. We have found the fit
\begin{equation}\nonumber
C_{FS} \left( n,l,m \right)= a_{lm} n^2 + b_{lm} n+c_{lm} 
\end{equation}
where the parameters a,b,c are given in Table 1 for two particular states with the corresponding correlation coefficient $R$ of the fit. It would be extremely interesting to show this result from Eqs. (\ref{Cfs2})-(\ref{Bnlm}) in a rigorous physico-mathematical way.
\begin{table}
\begin{tabular}{|c|c|c|c|c|}
 & $a$ & $b$ & $c$ & $R$ \\ 
\hline
$C_{FS} \left( n,0,0 \right)$ & \hspace*{1mm} 0.565\hspace*{1mm} & \hspace*{1mm}1.202\hspace*{1mm} & \hspace*{1mm}-1.270 \hspace*{1mm}&\hspace*{1mm} 0.999996\hspace*{1mm} \\ 
$C_{FS} \left( n,3,1 \right)$ & \hspace*{1mm}0.451\hspace*{1mm} &\hspace*{1mm} 0.459 \hspace*{1mm}&\hspace*{1mm} -4.672\hspace*{1mm} &\hspace*{1mm} 0.999998\hspace*{1mm}\\
\hline
\end{tabular}
\caption{Fisher-Shannon measure of the hydrogenic orbitals $(n,l,m)=(n,0,0)$ and $(n,3,1)$}
\end{table}

\section{Upper bounds to the Fisher-Shannon measure and shape complexity}

We have seen previously that, contrary to the Cramer-Rao measure whose expression can be calculated explicitly in terms of the quantum numbers $(n,l,m)$, the Fisher-Shannon measure and the shape complexity have not yet been explicitly found. This is basically because one of their two ingredients (namely, the Shannon entropic power) has not yet been computed directly in terms of the quantum numbers. Here we will calculate rigorous upper bounds to these two composite information-theoretic measures by means of the three quantum numbers of a generic hydrogenic orbital. Let us first gather the expressions
\begin{equation}\label{Cfs3}
C_{FS} \left[ \rho \right]= \frac{4 Z^2}{n^3} \left( n - \left| m \right| \right) \frac{1}{2 \pi e} e^{\frac{2}{3} S \left[ \rho \right]}
\end{equation}
for the Fisher-Shannon measure and
\begin{equation}\label{Csc3}
C_{SC} \left[ \rho \right]= Z^3 D(n,l,m) e^{S \left[ \rho \right]}
\end{equation}
for the shape complexity of the hydrogenic orbital $(n,l,m)$, where $S \left[ \rho \right]$ denotes the Shannon information entropy given by Eq. (\ref{expS}) and $D(n,l,m)$ has the exact value given by Eq. (\ref{Dnlm}). To write down these two expressions, we have taken into account Eqs. (\ref{Cfs}) and (\ref{Irho}) and Eqs. (\ref{expS}), (\ref{Csc}) and (\ref{Dnlm}), respectively. The exact calculation of the Shannon entropy $S \left[ \rho \right]$ is a formidable open task, not yet accomplished in spite of numerous efforts \cite{Y94,Y99,DMS}. Nevertheless, variational bounds to this information-theoretic quantity have been found \cite{A92, GAD, BIA} by means of one and two radial expectation values. In particular, for $\alpha=1$ we have that
\begin{equation}\label{cotaS}
S \left[\rho \right] \leqslant \ln \left[8 \pi \left( \frac{e \left\langle r \right\rangle }{3} \right)^3 \right]
\end{equation}

Then, taking into account that the expectation value $\left\langle r \right\rangle$ of the hydrogenic orbital $(n,l,m)$ is given \cite{D08, B57} by 
\begin{equation}\label{VarEspR}
\left\langle r \right\rangle = \frac{1}{2 Z} \left[ 3n^2- l(l+1) \right],
\end{equation}
so that
\begin{equation}\label{cotaS2}
S \left[\rho \right] \leqslant \ln \left\lbrace \frac{\pi e^3}{27 Z^3} \left[3 n^2 -l (l+1) \right]^3 \right\rbrace
\end{equation}

Now, from Eqs. (\ref{Cfs3}), (\ref{Csc3}) and (\ref{cotaS2}), we finally obtain the upper bounds 
\begin{equation}\label{cotaCfs}
C_{FS} \left[ \rho \right] \leqslant B_{FS} = \frac{2 e}{9 \pi^{1/3}} \frac{ n - \left| m \right| }{n^3} \left[ 3 n^2-l (l+1) \right]^2
\end{equation}
to the Fisher-Shannon measure, and
\begin{equation}\label{cotaCsc}
C_{SC} \left[ \rho \right] \leqslant B_{SC}= \frac{\pi e^3}{27} \left[ 3 n^2 - l(l+1) \right]^3 \times D(n,l,m) 
\end{equation}
to the shape complexity. It is worth noting that these two inequalities saturate at the ground state, having the values $\frac{2 e}{\pi^{1/3}}$ and $\frac{e^3}{8}$ for the Fisher-Shannon and shape complexity cases, respectively, when $n=1$, $l=0$, and $m=0$.

For the sake of completeness we plot in \textbf{Figure 6} and \textbf{Figure 7} the values of the ratios
\begin{figure}[h]
\includegraphics[scale=0.4,angle=270]{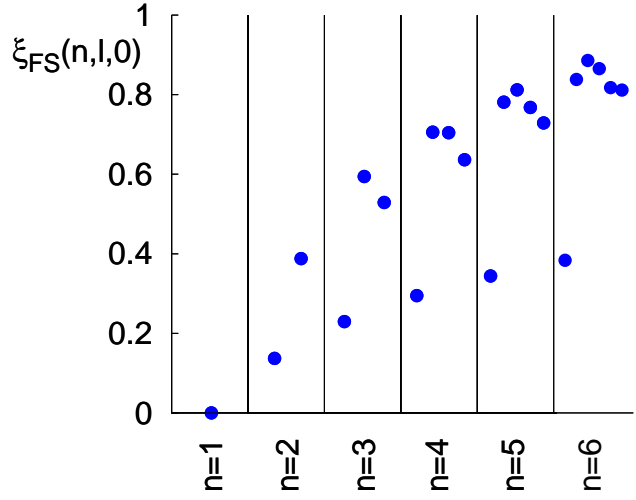}
\caption{Dependence of the Fisher-Shannon ratio, $\xi_{FS} (n,l,0)$, on the quantum numbers $n$ and $l$.}
\end{figure}

\begin{figure}[h]
\includegraphics[scale=0.4,angle=270]{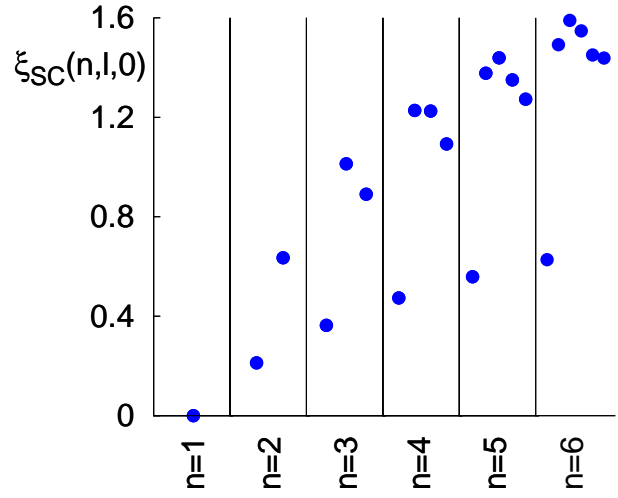}
\caption{Dependence of the shape-complexity ratio, $\xi_{SC} (n,l,0)$, on the quantum numbers $n$ and $l$.}
\end{figure}

\begin{equation}\nonumber
\xi_{FS} (n,l,m)= \frac{B_{FS}-C_{FS} \left[ \rho \right]}{C_{FS} \left[ \rho \right]}
\end{equation}
and
\begin{equation}\nonumber
\xi_{CS} (n,l,m)= \frac{B_{SC}-C_{SC} \left[ \rho \right]}{C_{SC} \left[ \rho \right]}
\end{equation}
for the Fisher-Shannon and the shape complexity measures, respectively, in the case $(n,l,m)$ for $n=1,2,3,4,5$ and $6$, and all allowed values of $l$. Various observations are apparent. First, the two ratios vanish when $n=1$ indicating the saturation of the inequalities (\ref{Csc3}) and (\ref{cotaCsc}) just mentioned. Second, for a manifold with fixed $n$ the greatest accuracy occurs for the states $s$. Moreover, the accuracy of the bounds decreases when $l$ is increasing up to the centroid of the manifold and then it decreases. Finally, the Fisher-Shannon bound is always more accurate than the Cramer-Rao bound for the same hydrogenic orbital.

\section{Conclusions}

In this work we investigate both analytically and numerically the internal disorder of a hydrogenic atom which gives rise to the great diversity and complexity of three-dimensional geometries for its configuration orbitals $(n,l,m)$. This is done by means of the following composite information-theoretic quantities: the Fisher-Shannon and the Cramer-Rao measures, and the shape complexity. The two former ones have a common ingredient of local character (the Fisher information) and a measure of global character, namely the Shannon entropic power in the Fisher-Shannon case and the variance in the Cramer-Rao case. The shape complexity is composed by two quantities of global character: the disequilibrium (whose explicit expression is here calculated for the first time in terms of the quantum numbers $n$, $l$ and $m$ of the orbital) and the Shannon entropic power.

We have studied the dependence of these three composite quantities in terms of the quantum numbers $n$, $l$ and $m$. It is found that:  (i) when $(l,m)$ are fixed, all of them have an increasing behaviour as a function of the principal quantum number $n$, with a rate of growth which is bigger in the Cramer-Rao and (even more emphatic) Fisher-Shannon cases; this is mainly because of the increasingly strong radial oscillating nature (when $n$ gets bigger), what is appropriately grasp by the Fisher ingredient of these two composite quantities; (ii) all of them decrease when the magnetic quantum number $\left| m  \right|$ is increasing, and the decreasing rate is much faster in the Cramer-Rao case and more emphatically in the Fisher-Shannon case; this is basically because of the increasingly weak angular oscillating nature when $\left| m  \right|$ decreases, what provokes the lowering of the Fisher ingredient of these two quantities; and (iii) all of them decrease when the orbital quantum number $l$ is increasing, and again, the decreasing rate is much faster in the Cramer-Rao and Fisher-Shannon cases; here, however, the physical interpretation is much more involved as it is duly explained in Section III.

Finally, we have used some variational bounds to the Shannon entropy to find sharp, saturating upper bounds to the Fisher-Shannon measure and to the shape complexity .

\section*{Appendix: Radial and angular contributions to the hydrogenic disequilibrium}

Here we briefly described the calculation of the radial and angular integrals involved in the calculation of the disequilibrium of the hydrogenic orbital $(n,l,m)$.

To calculate the radial integral $K(n,l,m)$ given by equation (\ref{Kn_rl}) we have use the orthonormal Laguerre polynomial ${\cal{L}}_{k}^{\alpha} (x)$, which has the relation
\begin{equation}\label{Lortonormal}
{\cal{L}}_{k}^{\alpha} (x)= \left[ \frac{\Gamma(k+\alpha+1)}{k!} \right]^{1/2} {\tilde{{\cal{L}}}}_{k}^{\alpha} (x)
\end{equation}
with the orthonormal Laguerre polynomial ${\tilde{{\cal{L}}}}_{k}^{\alpha} (x)$. Now, taking into account the linearization formula
\begin{equation}\label{LinearizacionL}
\left[ {\cal{L}}_{n_r}^{2l+1}  (x) \right]^2= \frac{\Gamma(2l+n_r+2)}{2^{2 n_r} n_r!} \sum_{k=0}^{n_r} \begin{pmatrix} 2n_r-2k \\[-1.5mm] n_r- k \end{pmatrix} \frac{(2 k)!}{k!} \frac{1}{\Gamma(2l+2+k)} {\cal{L}}_{2 k}^{4l+2}  (2x)
\end{equation}
and the orthogonality relation 
\begin{equation}\label{OrtogonalidadL}
\int_{0}^{\infty} \omega_\alpha (x) {\cal{L}}_{k}^{\alpha} (x) {\cal{L}}_{k^{'}}^{\alpha} (x)= \frac{\Gamma(k+\alpha+1)}{k!} \delta_{kk^{'}},
\end{equation}
one has
\begin{equation}\nonumber
\left\langle \rho \right\rangle_{R} = \frac{Z^3 2^{2-4n}}{n^5} \sum_{k=0}^{n_r} \begin{pmatrix} 2n_r-2k \\[-1.5mm] n_r- k \end{pmatrix}^2 \frac{(k+1)_k}{k!} \frac{ \Gamma \left(4l+2k+3 \right)}{ \Gamma^2 \left(2l+k+2 \right)}
\end{equation}

To calculate the angular integral $\left\langle \rho \right\rangle_{Y}$, we use the following linearization relation of the spherical harmonics
\begin{equation}\label{Ylm^2}
\left| Y_{lm} \left( \Omega \right)\right|^2= \sum_{l'=0}^{2l} \frac{{\hat{l}}^2 \hat{l'}}{\sqrt{4\pi}} \begin{pmatrix} l & l & l' \\[-1.5mm] 0 & 0 & 0 \end{pmatrix} \begin{pmatrix} l & l & l' \\[-1.5mm] m & m & -2m \end{pmatrix} Y_{l',2m}^{*} \left( \Omega \right),
\end{equation}
where $\hat{a}=\sqrt{2a+1}$ and the $3j$-symbols \cite{BIE} have been used. Then, taking into account Eqs. (\ref{DeseqRho}) and (\ref{Ylm^2}) one has that
\begin{equation}\label{DeseqY}
\left\langle \rho \right\rangle_Y = \sum_{l'=0}^{2l} \frac{{\hat{l}}^2 \hat{l'}}{\sqrt{4\pi}} \begin{pmatrix} l & l & l' \\[-1.5mm] 0 & 0 & 0 \end{pmatrix} \begin{pmatrix} l & l & l' \\[-1.5mm] m & m & -2m \end{pmatrix}  W\left( l,m \right),
\end{equation}
where $W \left(l,m \right)$ denotes the following integral of three spherical harmonics
\begin{equation}\label{int3Y}
W \left(l,m \right)=\int_{\Omega} Y_{l,-m} \left( \Omega \right)  Y_{l,-m} \left( \Omega \right)  Y^{*}_{l,2m} \left( \Omega \right) d\Omega 
\end{equation}

Moreover, taking into account the known general integral
\begin{equation}\nonumber
\int_{\Omega} Y_{l_1,m_1} \left( \Omega \right)  Y_{l_2,m_2} \left( \Omega \right)  Y^{*}_{l_3,m_3} \left( \Omega \right) d\Omega = \frac{\hat{l_1} \hat{l_2} \hat{l_3}}{\sqrt{4 \pi}} \begin{pmatrix} l_1 & l_2 & l_3 \\[-1.5mm] 0 & 0 & 0 \end{pmatrix} \begin{pmatrix} l_1 & l_2 & l_3 \\[-1.5mm] m_1 & m_2 & m_3 \end{pmatrix},
\end{equation} 
one has
\begin{equation}\nonumber
\left\langle \rho \right\rangle_{Y} = \sum_{l'=0}^{2l} \left( \frac{{\hat{l}}^2 \hat{l'}}{\sqrt{4\pi}} \right)^2 \begin{pmatrix} l & l & l' \\[-1.5mm] 0 & 0 & 0 \end{pmatrix}^2 \begin{pmatrix} l & l & l' \\[-1.5mm] m & m & -2m \end{pmatrix}^2
\end{equation}

\begin{acknowledgments}
The authors gratefully acknowledge the Spanish MICINN grant FIS2008-02380 and the grants FQM-1735 and 2445 of the Junta de Andaluc\'ia. They belong to the Andalusian research group FQM-207. S.L.R. and D.M. acknowledge the FPU and FPI scholarships of the Spanish Ministerio de Ciencia e Innovaci\'{o}n, respectively.
\end{acknowledgments}

\bibliographystyle{ppcf}
\bibliography{general}

\end{document}